# Contact Tracing Apps for COVID-19: Access Permission and User Adoption

Amal Awadalla Ali[1], Asma Hamid ElFadl[1], Maha Fawzy Abujazar[1], Sarah Aziz[1], Alaa Abd-Alrazaq[1], Zubair Shah[1], Samir Brahim Belhaouari[1], Mowafa Househ[1], Tanvir Alam[1,*]

[1] College of Science and Engineering, Hamad Bin Khalifa University, Doha, Qatar.

* Correspondence: talam@hbku.edu.qa

*Abstract—— Contact tracing apps are powerful software tools that can help control the spread of COVID-19. In this article, we evaluated 53 COVID-19 contact tracing apps found on the Google Play Store in terms of their usage, rating, access permission, and user privacy. For each app included in the study, we identified the country of origin, number of downloads, and access permissions to further understand the attributes and ratings of the apps. Our results show that contact tracing apps had low overall ratings and nearly 40% of the included apps were requesting "dangerous access permission" including access to storage, media files, and camera permissions. We also found that user adoption rates were inversely correlated to access permission requirements. To the best of our knowledge, our article summarizes the most extensive collection of contact tracing apps for COVID-19. We recommend that future contact tracing apps should be more transparent in permission requirements and should provide justification for permissions requested to preserve the app users' privacy.*

**Keywords— COVID-19; Coronavirus; Contact Tracing; Mobile Apps; Privacy**

## I. INTRODUCTION

The global outbreak of novel coronavirus disease 2019 (COVID-19) has caused unprecedented health, social, and economic effect on mankind [1]. The COVID-19 disease, caused by novel coronavirus (SARS-CoV-2), has spread rapidly from person to person (i.e., being in close contact, through respiratory droplets from coughing, sneezing, or touching nose or eyes after touching infected objects, or other media) around the world which mainly caused severe respiratory illness to the infected person [2]. Considering the severity of this disease, The World Health Organization (WHO) declared the COVID-19 outbreak as a pandemic on March 12, 2020 [3]. As of August 07, 2020, more than 19 million people have been infected, and nearly 720000 fatalities resulted from COVID-19 [4].

For the containment of the COVID-19, the WHO recommended people to practice simple precautions like hand washing, covering the mouth and nose by wearing appropriate masks, practicing physical distancing of one meter (three feet), and avoid large group gatherings [5]. At the same time, tremendous effort from the scientific community has been invested in discovering treatments, including antiviral drugs, and vaccines. Although no effective medications or vaccines have been discovered so far, to eradicate the COVID-19, different pharmaceutical prophylaxis and treatment such as Dexamethasone, Remdesivir, and few others are tested for their efficacy [6]. Many other attempts are underway to discover an effective vaccine, as well [7]. As of July 2020, more than 25 candidate vaccines for COVID-19 are at different phases of clinical trials [8]. In fact, the vaccine development is a lengthy, laborious, and expensive process [9], and the whole world is waiting for an effective vaccine to come out to eradicate this disease.

With the above backdrop and the lack of effective therapeutic solutions against COVID-19, contact tracing mobile apps could be considered as a viable option to control or minimize the spread of the COVID-19. These days usage of mobile technologies has grown rapidly with the increasing usage of mobile electronic devices such as smartphones. As part of controlling the spread of COVID-19, contract tracing apps for smart mobile phones are already implemented in different countries [10]. Those apps are defined as mobile software applications that use digital contact tracing in response to the COVID-19 pandemic. These apps are already rolled out to identify people who were in close contact with other infected persons, to alert them to step away, daily updates on the census for the COVID-19 cases according to their locations, and necessary safeguarding instructions from the government [11].

As the number of contact tracing apps is surging, it is essential to understand its usage, ratings, permission, and privacy setting to understand users' adoption. Lalmuanawma et al. summarize the contribution of artificial intelligence (AI) in combating COVID-19 [12] and mentioned a list of 36

TABLE I. EXISTING LITERATURES ON THE REVIEW OF COVID-19 CONTACT TRACING APPS

| Publication Date and Reference | Types of apps Covered | Summary | Number of Apps mentioned/ Analyzed |
|---|---|---|---|
| March, 2020 [12] | All apps related to COVID-19 | Country and location tracking system (GSM, Bluetooth) were mentioned | 36 |
| May, 2020 [13] | Only contact tracing apps | Country name, App name, the number of users, and the underpinning technologies (GPS, QR codes, Bluetooth) were mentioned | 10 |
| July, 2020 [14] | Only contact tracing apps | Architecture type (Centralized, Decentralized and Hybrid) and Possible Attacks on Tracing Apps and Protocols | 15 |
| July, 2020 [15] | Only contact tracing apps | Applications, Country, Downloads, and Version were mentioned | 34 |
| July, 2020 [17] | Only contact tracing apps | App, Platform, Permissions Requested, Privacy Policy, Country, No. of Downloads, TLS/SSL, App Reviews were mentioned | 23 |

COVID-19 related apps. Li and Guo analyzed the flaw and data privacy issue of 10 contact-tracing apps [13]. Ahmed et al. analyzed 15 contact tracing apps covering their system architecture, location estimation system, data management, privacy, and security vulnerability [14]. Sun et al. summarized

34 contact tracing apps considering several aspects of privacy preservation for the uses. The authors also described the design paradigm, the potential vulnerability of private data, the robustness of privacy protection schemes for the apps [15]. Azad et al. analyzed 23 contact tracing apps considering their security, permission, and privacy [16]. The authors also mentioned that some contact tracing apps failed to follow proper security measures while exchanging data to and from the data centers. Table I summarizes a list of existing review articles that summarize contact tracing apps for COVID-19 from different perspectives.

But the existing review articles on COVID-19 contact tracing apps covered either too few numbers of contract tracing apps, or they considered only privacy aspects of the apps without describing other relevant aspects like ratings, downloads, and permission details. This study aims to analyze the existing COVID-19 contact tracing apps in terms of their usage, rating, access permission, and user privacy. To the best of our knowledge, this article covers the largest number of contact tracing apps for COVID-19.

## II. MATERIALS AND METHODS

### A. Collection of Contact Tracing Apps

To collect a list of COVID-19 contract tracing apps, we searched for apps that have been officially released by different countries. We searched for existing literature in PubMed and Google Scholar. Some contact tracing apps (e.g., Ehteraz, Covi-ID, and many others) are released into Google Play Store without corresponding research articles published; therefore, for those contact tracing apps, we also searched the Google Play Store. As of August 10, 2020, we were able to collect a list of 53 contact tracing apps from 53 countries (Table II).

### B. Metadata Extraction for the Selected Contact Tracing Apps

We collected the metadata for each contact tracing app from Google Play Store, such as a short description, categories, average rating, number of users who rated the app, last update version, size, number of installations, current version, required android version, access permissions, developed by, privacy policy website, and application website from Google Play Store. Supplementary Table 1 provides the details of this metadata for each app.

### C. Mapping of Google Play Store Permissions into Android Dangerous Permissions

Google provides a list of permissions corresponding to individual apps, with a purpose to protect the privacy of the user. These permissions are divided into different subcategories. To determine the protection level of these android apps, we considered the Android permission levels [17] and mapped them to Google Play Store permissions available from the app metadata. Owing to the fact that we did not use the source codes of android apps, thus, to overcome this issue between manually gathering of permissions rather than source code analysis, we mapped the dangerous android permissions as mentioned in documentation to Google Play permissions considering that they mean same according to descriptions provided by each one. These mapping could be observed in Supplementary Table 2.

### D. Correlation analysis between app permissions and ratings and download

Correlation analysis, along with statistical significance, of apps permission and corresponding ratings and normalized number of downloads considering the population size of respective countries were computed using the SPSS software.

## III. RESULTS

### A. Category of Contact Tracing Apps

The list of contact tracing apps mainly belonged to four categories: (a) health & fitness (n=25), (b) medical (n=22), (c) lifestyle (n=3), and (d) social (n=2) (Fig. 1). We found only one contact tracing app MorChana, from Thailand, which was under the category of the tool. From Fig. 1, we noticed that the lifestyle group had the highest average rating equal to 3.8, followed by the health and fitness group with an average rating equal to 3.7, then medical group with an average rating equal to 3.6, and the social group with an average rating equal to 3.5.

### B. Download statistics of Contact Tracing Apps

Fig. 2 illustrates the rating of contact tracing apps and the number of downloads for the apps. We observed that the contact tracing apps in the health & fitness category have the highest number of downloads among all apps. Interestingly we found that the Aarogya Setu app developed for India had the highest number of downloads by more than 100 million users, followed by Coron App developed for Colombia having more than 10 million downloads. Corona Warn App, Hayat Eve Sigar, CUIDAR COVID-19 developed for Germany, Turkey, and Argentina respectively were downloaded by more than 5 million.

On the other hand, the lowest number of app installs was Corona 360 developed for South Korea had only 100 downloads, followed by Covi-ID developed for South Africa having 500 downloaded. CovTracer, GH Covid-19 Tracker App, Best Covid Gibraltar, and Kenya Covid-19 Tracker developed for Cyprus, Ghana, Gibraltar and Kenya, respectively had over 1000 plus downloads only.

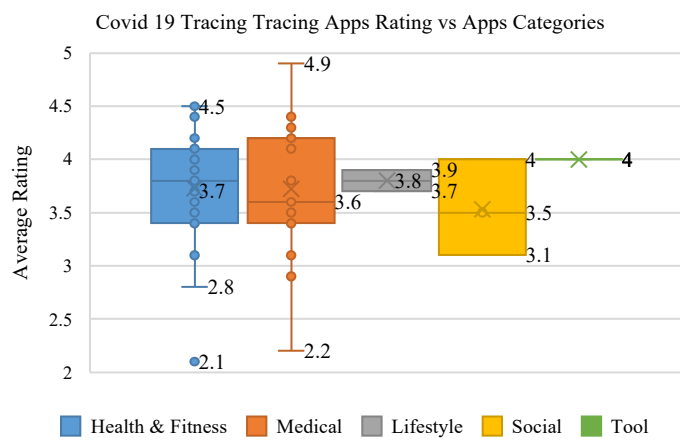

Fig. 1. The average rating of COVID-19 contact tracing apps in each category.

TABLE II. FEATURES OF 53 CONTACT TRACING APPLICATIONS

| No. | Country | Contact tracing App | Categories | Updated | Size | Current Version |
|---|---|---|---|---|---|---|
| 1 | Algeria | Coronavirus Algérie | Health & Fitness | 23-Mar-20 | 9.5M | 1.0.3 |
| 2 | Argentina | CUIDAR COVID-19 ARGENTINA | Medical | 19-Jul-20 | 16M | 3.5.0 |
| 3 | Australia | COVIDSafe | Health & Fitness | 4-Jul-20 | 11M | 1.0.33 |
| 4 | Austria | Stopp Corona | Medical | 28-Jun-20 | 5.3M | 2.0.3.1057-QA_245 |
| 5 | Bahrain | BeAware | Health & Fitness | 5-Jul-20 | 7.0M | 0.1.9 |
| 6 | Brunei | BruHealth | Health & Fitness | 15-Jul-20 | 16M | 1.1.13 |
| 7 | Bulgaria | ViruSafe | Health & Fitness | 27-May-20 | 3.7M | 1.0.3 |
| 8 | Canada | Canada COVID-19 | Medical | 10-July-20 | 8.3M | 4.0.0 |
| 9 | Colombia | CoronApp | Health & Fitness | 1-Jul-20 | 9.9M | 1.2.50 |
| 10 | Cyprus | CovTracer | Health & Fitness | 11-Jun-2020 | 28M | 2.0.2 |
| 11 | Czech Republic | eRouška (eFacemask) | Health & Fitness | 22-May-2020 | 3.7M | 1.0.437 |
| 12 | Denmark | Smittestop | Medical | 9-Jul-2020 | 96M | 1.0.3 |
| 13 | Fiji | CareFiji | Medical | 23-Jun-2020 | 8.7M | 1.0.46 |
| 14 | France | StopCovid | Medical | 26-Jun-2020 | 17M | 1.1.0 |
| 15 | Germany | Corona-Warn-App | Health & Fitness | 7-Jul-2020 | 31M | 1.0.5 |
| 16 | Ghana | GH Covid-19 Tracker App | Health & Fitness | 8-May-2020 | 4.0M | 1.0.3 |
| 17 | Gibraltar | Beat Covid Gibraltar* | Medical | 25-Jun-2020 | 114M | 1.0.1.18 |
| 18 | Hong Kong | StayHomeSafe | Health & Fitness | 22-Jun-2020 | 11M | 0.8.4 |
| 19 | Hungary | VírusRadar | Social | 15-May-2020 | 12M | 1.0.0 |
| 20 | Iceland | Rakning C-19 | Lifestyle | 14-Jul-2020 | 28M | 2.1.1 |
| 21 | India | Aarogya Setu | Health & Fitness | 8-Jul-2020 | 4.2M | 1.4.1 |
| 22 | Indonesia | PeduliLindungi | Medical | 25-Jun-2020 | 7.2M | 2.2.2 |
| 23 | Ireland | HSE Covid-19 App | Medical | 25-Jun-2020 | 114M | 1.0.40 |
| 24 | Israel | HaMagen | Health & Fitness | 16-Jun-2020 | 26M | 1.4.7 |
| 25 | Italy | Immuni | Medical | 10-Jul-2020 | 28M | 1.3.0 |
| 26 | Japan | COCOA | Medical | 14-Jul-2020 | 100M | 1.1.2 |
| 27 | Jordan | AMAN App - Jordan | Medical | 24-May-2020 | 3.8M | 1 |
| 28 | Kenya | Kenya Covid-19 Tracker | Medical | 28-Apr-2020 | Varies with device | v0.4.38 |
| 29 | Kuwait | Shlonik | Health & Fitness | 21-Jun-2020 | Varies with device | Varies with device |
| 30 | Latvia | Apturi Covid | Medical | 26-Jun-2020 | 9.3M | 1.0.47 |
| 31 | Malaysia | MyTrace | Lifestyle | 26-Apr-2020 | 30M | 1.0.30 |
| 32 | Mexico | CovidRadar | Health & Fitness | 2-May-2020 | 2.3M | 1.0.2 |
| 33 | Morocco | wiqaytna | Medical | 19-Jun-2020 | 5.9M | 1.1.0 |
| 34 | New Zealand | NZ COVID Tracer | Health & Fitness | 20-Jun-2020 | 16M | 1.1.1 |
| 35 | North Macedonia | StopKorona | Social | 7-May-2020 | 15M | 1.1.0 |
| 36 | Northern Ireland | COVID-19 NI | Medical | 1-Jun-2020 | 54M | 1.7 |
| 37 | Norway | Smittestopp | Medical | 8-Jun-2020 | 4.2M | 1.3.0 |
| 38 | Philippines | StaySafe PH | Medical | 6-Jul-2020 | 14M | 1.0.3 |
| 39 | Poland | ProteGO | Medical | 6-Jul-2020 | 13M | 4.2.1 |
| 40 | Portugal | Estamos ON - Covid19 | Social | 28-Mar-2020 | 8.5M | 1.0.1 |
| 41 | Qatar | Ehteraz | Health & Fitness | 2-Jul-2020 | 34M | 8.0.4 |
| 42 | Saudi Arabia | Tabaud | Health & Fitness | 10-Jul-2020 | 6.2M | 1.1.0 |
| 43 | Singapore | TraceTogether | Medical | 7-Jul-2020 | 26M | 2.1.4 |
| 44 | South Africa | Covi-ID | Health & Fitness | 5-Jun-2020 | 27M | 1 |
| 45 | South Korea | Corona 360 | Medical | 18-Apr-2020 | 4.3M | 2.2.2 |
| 46 | Switzerland | SwissCovid | Health & Fitness | 20-Jun-2020 | 12M | 1.0.5 |
| 47 | Thailand | MorChana | Tool | 6-Jun-2020 | 47M | 1.8.4 |
| 48 | Tunisia | E7mi | Health & Fitness | 22-May-2020 | 11M | 1.1 |
| 49 | Turkey | Hayat Eve Sigar | Health & Fitness | 24-Jun-2020 | 6.0M | 2.0.9 |
| 50 | UAE | TraceCovid | Medical | 13-Apr-2020 | 10M | 1.1.6 |
| 51 | Uruguay | Coronavirus UY | Health & Fitness | 1-Jul-2020 | 23M | 4.3.9 |
| 52 | USA - State of Utah | Healthy Together - COVID-19 | Health & Fitness | 15-Jul-2020 | 42M | 1.2.48 |
| 53 | Vietnam | BlueZone | Health & Fitness | 15-May-2020 | 7.4M | 2.0.2 |

The normalized number of downloads for BruHealth app developed for Brunei (23.1%), Coron App developed for Colombia (19.9%), and TraceTogether developed for Singapore (17.5%) were higher compared to other apps. The normalized number of downloads for Corona 360 developed for South Korea (0.0002%), Covi-ID developed for South Africa (0.0009%), and Kenya Covid-19 Tracker developed for Kenya (0.0019%) were lower compared to other apps.

*C. App Permissions and Android Versions*

App permissions in Google Play Store have been categorized according to the android version. For Android 5.1 and lower, there are 16 permissions groups, while for android 6.0 and higher, there are nine permissions groups (Supplementary Table 2). Some permissions are not categorized in the permissions group, so we categorized them as "other" [18]. Furthermore, users who use Android 6.0 and higher can control which permission group can access device information [19]. After the user downloads an application from Google Play Store by using Android 6.0 and higher each permission group will be shown separately, and the user can control which permission can access the device information [19]. Whereas in Android 5.1 and lower, a list of permissions groups are shown before downloading the application, and the user can accept all the permissions groups and download the app or decline it and the app will not be downloaded [18]. From the pool of 53 contact tracing apps, we found that 27 apps are suitable for Android version 5.0 and higher (Fig. 3).

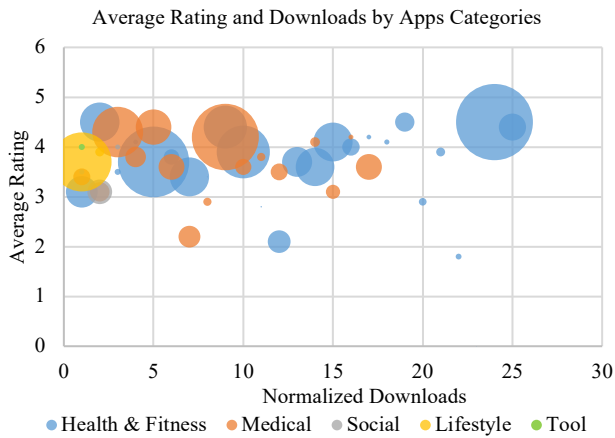

Fig. 1. Y-axis represents the apps' average rating, and the X-axis represents the normalized number of downloads. As the number of populations differs from country to country, we divided the number of actual downloads by the total population to normalize the value in X-axis.

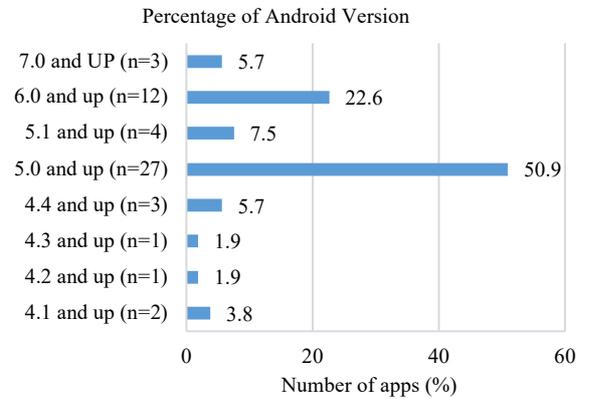

Fig. 2. COVID-19 contact tracing apps and corresponding Android version.

*D. Overview of Access Permissions for Contact Tracing Apps*

Fig. 4 summarizes the access permission requested by the 53 contact tracings apps. As shown in Fig. 4, all contact tracing apps require full network access permission. Most apps require access to view network connection (n=52) and prevent devices from sleep (n=46). About three-quarters of apps require access to run at startup (n=40) and pair with Bluetooth devices (n=40). Approximately 71.1% of apps require access to the location (n=38) and receive data from the internet (n=38). Nearly half of the apps require Bluetooth access (n=28). More than 30% contact tracing apps require access to storage (n=20), photo/media/file (n=20) and camera (n=16).

The least frequently access permissions were control vibration (n=14), Wi-Fi connection information (n=13), device and app history (n=7), phone (n=6), device ID and call information (n=4), draw over other apps (n=4), Google Play license check (n=4), contacts (n=3), change network connectivity (n=3), connect and disconnect from Wi-Fi (n=3), read Google service configuration (n=3), microphone (n=2), create accounts and set passwords (n=2), read sync settings (n=2), toggle sync on and off (n=2). Only 1.9% of apps require to access change your audio settings (n=1), disable your screen lock (n=1), send sticky broadcast (n=1), set an alarm (n=1), control flashlight (n=1), uninstall shortcuts (n=1), install shortcuts (n=1), and calendar (n=1).

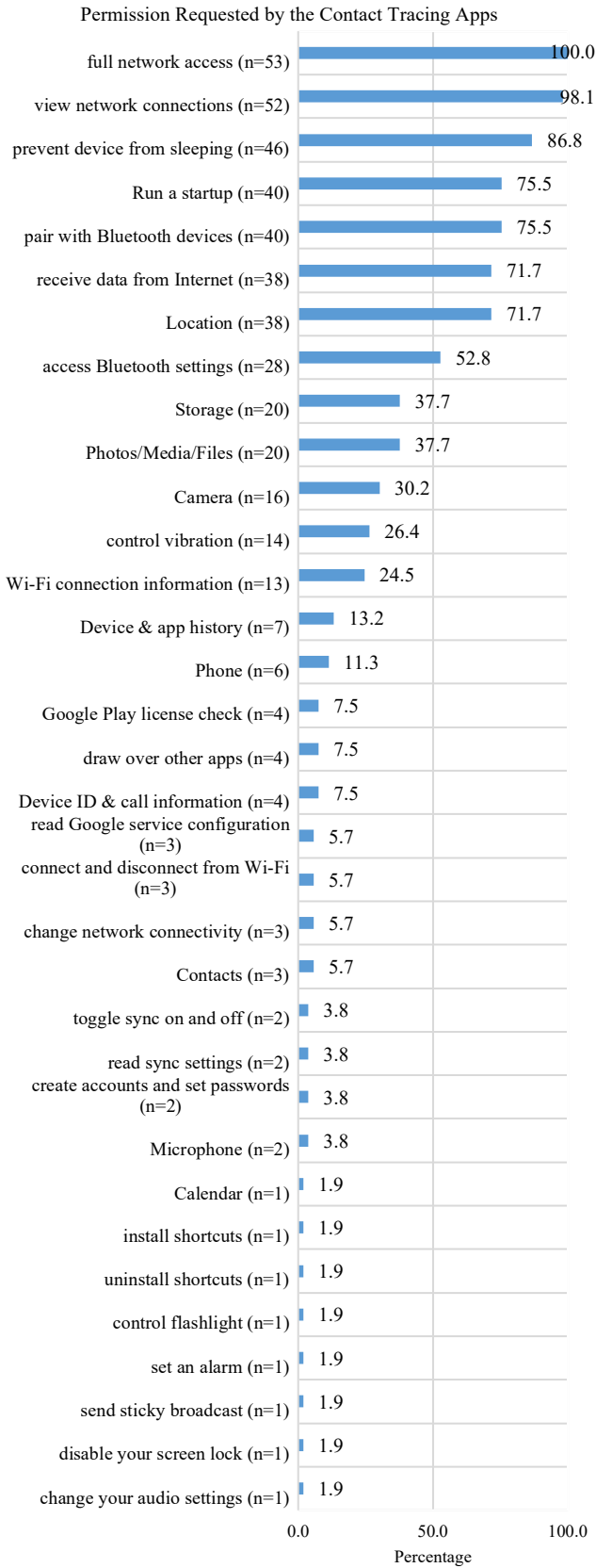

Fig. 3. Summary of permission requested by the contact tracing apps

*E. Dangerous Permissions*

Dangerous permissions are referred to as accesses made by apps to use data and resources that are private, store data in devices or operations related to other apps [20]. For example, the ability to read the user's contacts is dangerous permission as the app tries to get control over the private data. These dangerous permissions withhold great importance as they could lead to potential privacy-data breaches [21]. Table III shows the percentage of potentially dangerous permission requests [17] made by contact tracing apps. For the purpose of COVID-19 contact tracing, the location permission is of great importance as the app could alert a user when he/she is in close contact with an infected person. But more than 30% of contact tracing apps require read/write access to USB storage and multimedia (photo/media/file), which may raise a great concern for the app users.

TABLE III. NUMBER AND PERCENTAGE OF APPS THAT REQUEST DANGEROUS PERMISSIONS

| Google Play Store Permissions | Android Dangerous Permissions | Number | Percentage |
|---|---|---|---|
| Location | precise location (GPS and network-based) | access fine location | 38 | 71.7 |
| | approximate location (network-based) | access coarse location | 29 | 54.7 |
| Contacts | read your contacts | read contacts | 3 | 5.7 |
| Calendar | read calendar events plus confidential information | write calendar | 1 | 1.9 |
| | add or modify calendar events and send email to guests without owners' knowledge | read calendar | 1 | 1.9 |
| Phone | directly call phone numbers | read call log | 3 | 5.7 |
| | read phone status and identity | read phone state | 5 | 9.4 |
| Photos/Media/Files | Read the contents of your USB storage | read external storage | 20 | 37.7 |
| | Modify or delete the contents of your USB storage | write external storage | 18 | 34.0 |
| Camera | take pictures and videos. | camera | 16 | 30.2 |
| Microphone | record audio | record audio | 2 | 3.8 |
| Storage | read the contents of your USB storage | read external storage | 20 | 37.7 |
| | modify or delete the contents of your USB storage | write external storage | 18 | 34.0 |
| Device ID & call information | read phone status and identity) | read phone state | 4 | 7.5 |

## F. Correlation analysis between contact tracing apps permission and ratings and downloads

Correlation between the app's permission and corresponding rating and downloads were computed (Table 4). From Table IV, we can observe that both ratings and downloads were inversely related to the dangerous permissions like location, storage, photos/media/files. The contact tracing apps with dangerous permission, storage had lower downloads (correlation: -0.18; p-value:0.11). The number of downloads is lower for the apps having photos/media/files permission (correlation: -0.18; p-value:0.11). A similar trend was observed for the apps requiring wi-fi connection information (correlation: -0.10; p-value:0.24), though the correlation was not statistically significant.

TABLE IV. CORRELATION BETWEEN APPS PERMISSION AND RATING AND DOWNLOADS. ROWS IN THE TABLE ARE SORTED IN DESCENDING ORDER OF CORRELATION BETWEEN PERMISSION AND DOWNLOAD. STD_DOWNLOAD: STANDARDIZED DOWNLOAD. *: DANGEROUS PERMISSION.

| Permission Type | Correlation (Rating) | P-value (Rating) | Correlation (std_download) | P-value (std_download) |
|---|---|---|---|---|
| Phone* | 0.04 | 0.38 | 0.29 | **0.02** |
| pair with Bluetooth devices | 0.16 | 0.13 | 0.26 | **0.03** |
| Change network connectivity | 0.25 | 0.04 | 0.24 | **0.04** |
| control flashlight | 0.22 | 0.06 | 0.24 | **0.04** |
| set an alarm | -0.01 | 0.47 | 0.23 | **0.05** |
| view network connections | -0.14 | 0.16 | 0.18 | 0.10 |
| Run a startup | 0.06 | 0.35 | 0.18 | 0.10 |
| Contacts* | 0.08 | 0.29 | 0.17 | 0.11 |
| Calendar* | 0.22 | 0.06 | 0.16 | 0.12 |
| prevent device from sleeping | 0.08 | 0.30 | 0.12 | 0.19 |
| Camera* | 0.06 | 0.34 | 0.11 | 0.22 |
| disable your screen lock | -0.01 | 0.47 | 0.11 | 0.22 |
| connect and disconnect from Wi-Fi | 0.00 | 0.49 | 0.11 | 0.22 |
| Device ID & call information* | -0.06 | 0.35 | 0.11 | 0.23 |
| read Google service configuration | 0.06 | 0.33 | 0.07 | 0.30 |
| Microphone record audio* | 0.20 | 0.08 | 0.06 | 0.34 |
| access Bluetooth settings | 0.17 | 0.11 | 0.05 | 0.37 |
| Device & app history | 0.20 | 0.08 | 0.04 | 0.40 |
| create accounts and set passwords | 0.06 | 0.33 | 0.01 | 0.46 |
| read sync settings | 0.06 | 0.33 | 0.01 | 0.46 |
| toggle sync on and off | 0.06 | 0.33 | 0.01 | 0.46 |
| Google Play license check | 0.22 | 0.06 | 0.00 | 0.49 |
| change your audio settings | 0.10 | 0.25 | 0.00 | 0.49 |
| send sticky broadcast | 0.10 | 0.25 | 0.00 | 0.49 |
| receive data from Internet | 0.41 | 0.00 | -0.01 | 0.48 |
| control vibration | 0.12 | 0.20 | -0.02 | 0.44 |
| draw over other apps | -0.15 | 0.14 | -0.02 | 0.43 |
| Wi-Fi connection information | 0.09 | 0.27 | -0.10 | 0.24 |
| Location* | -0.07 | 0.32 | -0.11 | 0.22 |
| Photos/Media/Files* | 0.15 | 0.14 | -0.18 | 0.11 |
| Storage* | 0.15 | 0.14 | -0.18 | 0.11 |
| install shortcuts | -0.23 | **0.05** | -0.22 | **0.06** |
| uninstall shortcuts | -0.23 | **0.05** | -0.22 | **0.06** |

## IV. DISCUSSIONS

This article aimed to provide an overview of the COVID-19 contact tracing apps, including the permissions mentioned in the Google Play Store for the corresponding apps. Based on our analysis, we found that the average ratings for the contact tracing apps are not very high. The highest average rating for contact tracing apps was 3.8; this indicates the existence of avenues that need to be addressed by the developers and policymakers to improve the user adoption and rating for the apps. Most of the contact tracing apps were supporting Android version 5.0 and above. If the apps were compatible with the older version of Android as well, then we could expect more downloads as some of the Android users may not update their system for a long time.

The Number of downloads was varying for the contact tracing apps from country to country. On the higher side, we noticed 5-100 million downloads, and on the lower end, we observed only a few hundred downloads. Additionally, smartphone penetration is low in lower-to-middle income countries (e.g., Bangladesh, India, Pakistan), which may disenfranchise a certain group of population in lower-to-middle income countries. For the containment of COVID-19, successful outreach for contact tracing apps is mandatory; otherwise, a certain group of the population might be left without the app, causing a different rate of infection. So, we would recommend investigating the driving force for the installation of contact tracing apps across different countries.

Interestingly, some contact tracing apps were requiring access to set the alarm, flashlight, set shortcut which was entirely unnecessary for contact tracing. Also, apps that require access to photos/media/files can read sensitive log data, web bookmarks, and history or retrieve internal system state and running apps [18], [22]. As a result, an inverse relationship between the dangerous access permission of the apps and user adoption (number of downloads and rating) was observed.

Privacy and mobile health are interconnected because the information is shared and accessed using different means. Tracing apps may obtain personal information of people such as location, phones, camera, microphone, contacts, storage, and many others, which may lead to the breach of privacy. Thus, security and privacy policies should be explained clearly for all the apps in a simple language for users. Relevantly, Amnesty International UK pointed out seven principles at the plan of the UK Government's decision to roll-out a contacting-tracing app [23]. Therefore, governments should focus on contact tracing apps that are privacy-preserving following an acceptable guideline [23], and that would gain the confidence of the citizens. Overriding the consent and privacy of people for the sake of COVID-19 surveillance may raise the distrust in its population and, ultimately, lead to poorer health outcomes in this COVID-19 pandemic.

## V. CONCLUSION AND FUTURE WORKS

In this article, we have summarized 53 contact tracing apps developed to combat COVID-19 highlighting their ratings, user adoption, and access permissions. Some countries declared the ordinance for the usage of contact tracing apps as mandatory (e.g., China, Qatar), while other countries kept it as optional. Mandatory usage of apps may raise the eyebrow in the community considering the surveillance and privacy issues [24]. On the other hand, opt-in contact tracing initiatives require a lot of motivation and public awareness for the users. We believe our study will provide the governments, policymakers, and apps development industry an overall picture of the usage, ratings, and privacy aspects of these contact tracing apps. Further investigation on these features will enable us to develop a secure and privacy-preserving user-friendly contact tracing app for the containment of COVID-19. We only considered contact tracking apps from Google Play Store. In future, we will extend the list of contact tracing apps as well as include apps from the Apple App Store.

## SUPPLEMENTARY FILES

Supplementary files for this article are accessible in github: https://github.com/tanviralambd/ContactTracing.